# Longitudinal dynamic modelling and control for a quad-tilt rotor UAV

William Smith, Xinhua Wang
Aerospace Engineering,
University of Nottingham, UK
Email: wangxinhua04@gmail.com

**ABSTRACT**
Tilt rotor aircraft combine the benefits of both helicopters and fixed wing aircraft, this makes them popular for a variety of applications, including Search and Rescue and VVIP transport. However, due to the multiple flight modes, significant challenges with regards to the control system design are experienced. The main challenges with VTOL aircraft, comes during the dynamic phase (mode transition), where the aircraft transitions from a hover state to full forwards flight. In this transition phase the aerodynamic lift and torque generated by the wing/control surfaces increases and as such, the rotor thrust, and the tilt rate must be carefully considered, such that the height and attitude remain invariant during the mode transition.
In this paper, a digital PID controller with the applicable digital filter and data hold functions is designed so that a successful mode transition between hover and forwards flight can be ascertained. Finally, the presented control system for the tilt-rotor UAV is demonstrated through simulations by using the MATLAB software suite. The performance obtained from the simulations confirm the success of the implemented methods, with full stability in all three degrees of freedom being demonstrated.

## 1 INTRODUCTION

### 1.1 Tilt Rotor Aircraft and UAVs

Modern air travel has largely been achieved by two different aircraft types, aeroplanes and helicopters. Both of these have various advantages and disadvantages, for example, helicopters have a limited top speed and endurance due to the retreating blade stall phenomena and the fact that they make use of powered lift [1]. Similarly, the disadvantage of a conventional aeroplane is the significant length of runway required for the take-off velocities to be achieved. A VTOL (Vertical Take-Off and Landing) aeroplane can eliminate these drawbacks. This is achieved by taking off vertically like a helicopter, vectoring its thrust and transitioning into an aeroplane.
Being able to transition from helicopter-like flight to full aeroplane-like flight, means small space requirements and high forward flight speeds and endurance can be achieved.

### 1.2 Control Challenges

The ability of tilt rotors to transition from one aircraft type to another leads to drastically different dynamical behaviour between the hover and forward flight regimes. The arrangement of the aircraft type leads to unstable dynamic behaviour, the behaviour is sufficiently unstable that it is expected that even a competent pilot would struggle to control the individual rotor thrust with enough fidelity to maintain stable flight. It is because of these reasons that a controller is required to be introduced. Doing so will allow the individual rotor thrust and/or control surfaces to be augmented (with sufficient fidelity) such that the stability of the aircraft can be maintained [2].

With the aircraft implementation described in section 3, the moveable free wings cause a significant pitching moment when they are rotated from the vertical to the horizontal position as such they present a significant challenge when it comes to controlling the aircraft in the transition stage. A strategy for minimising the effect of these free wings is therefore discussed later in this paper.

### 1.3 Flight Mode Conversion

In order to transition between the helicopter mode and aeroplane mode, the rotors must tilt from the vertical position to the horizontal one [3]. During this tilting process, the thrust generated by the rotors is partially redirected into the horizontal direction, this makes the aircraft accelerate. During this acceleration phase, the velocity of the aircraft naturally increases and as such, the amount of lift generated by the wing also increases. Conversion between the flight modes can be considered successful if the growth of lift generated by the wing matches that of the reduction in rotor lift [3]. This, therefore, means that the rate at which the rotors are tilted must be carefully selected.

### 1.4 Conversion Corridor

The conversion corridor (also known as the transition corridor), outlines the safe conversion from helicopter flight to aeroplane flight, an example of which is shown in Figure 1. The left-hand side of the diagram outlines the minimum airspeed and rotor (nacelle) angle [3]. If the rotors are tilted too quickly, there is a risk that the left-hand limit (wing stall), could be violated and as such, it is possible that the summation of the wing lift and rotor thrust is insufficient to maintain altitude [3]. Furthermore, the right-hand side limit is a function of power, control and fatigue limits due to unsteady rotor loads [3]. Rotating the rotors too slowly can lead to these limits being breached [3] - calculating these limits, however, is outside the scope of this paper.

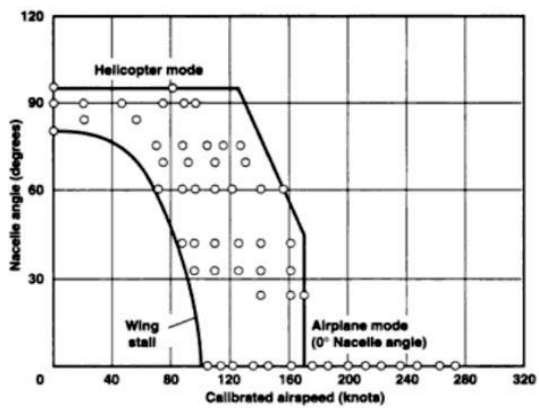

**Figure 1 - Conversion Corridor for the XV-15 [4]**

## 1.5 Rotor Tilt Rate

Numerous organisations have conducted ventures into tilt-rotor aircraft, analysis of their mode transition protocols will form the basis of the initial design. For the XV-15, the normal rotor tilt rate was $7.5°s^{-1}$ [3] This, therefore, allowed a complete mode transition in 12 seconds [3]. Similarly, the Leonardo AW609 has a rotor tilt rate of $3°s^{-1}$, or $8°s^{-1}$, depending on the current rotor tilt angle [3]. For both of these aircraft, the pilot can pause tilting at any point during the mode transition [3] and increase or decrease the airspeed in line with the restrictions set out by the conversion corridor [3]. This, therefore, means the controller must be designed such that movement within this conversion corridor is possible.

## 2 MODELLING AND DIGITAL CONTROL IMPLEMENTATION

Traditionally, analogue control has been used to implement aerospace control systems. Such systems use the Laplace/frequency domain to build the controller (and simulate any applicable models). However, with advances in low-cost digital computing, it has become possible to implement digital controllers. By replacing hardware control with a software implementation, many advantages can be realised. One such advantage is the increased flexibility and therefore reduction in cost for the initial design phase and any potential midlife updates.

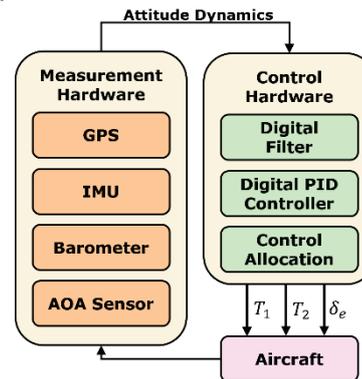

**Figure 2 - Implementation in a real aircraft**

When a digital control system is implemented in an aircraft, multiple sensors must be used to ascertain the current state of the aircraft. To highlight this, Figure 2 shows a schema for this conceptual control system. The measurement equipment such as IMU (Inertial Measurement Units, Barometers, GPS (Global Positioning Systems) and AoA (Angle of Attack) allow for the position, velocities and accelerations in all degrees of freedom to be ascertained. This data constructs the positional/attitude dynamics of the aircraft and therefore forms the inputs to the control system.

## 2.1 Finite Difference Method for digital modelling and control

In general, a continuous (analogue) system will be described by a series of differential equations. It is not possible to directly use such signals in the digital domain. Therefore, to implement the digital model/controller, the differential equations must be linearised, doing so leads to linear difference equations [5]. By using discrete data, it is possible to estimate the current state of the system at any given time. Furthermore, due to the linearization of the system, the output of the system is directly affected by the input to the loop, therefore meaning the input of the k[th] loop is produced by the output of the (k-1)[th] loop.

### 2.1.1 Data Hold Function for analogue to digital converter

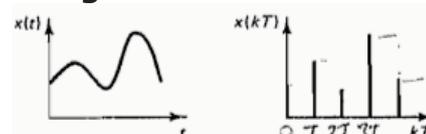

**Figure 3 - Sampling of a continuous system [5]**

As mentioned, it is not possible to directly use an analogue signal in the digital domain.

Therefore, one must use an analogue to digital converter to convert the analogue signal into a digital form.

As shown in Figure 3, a signal sampled at a sample time $T$, will effectively form a series of pulses [5]. In order to make the pulsed signal useable in the digital domain, a continuous signal must be approximated, this can be done using a data hold function.

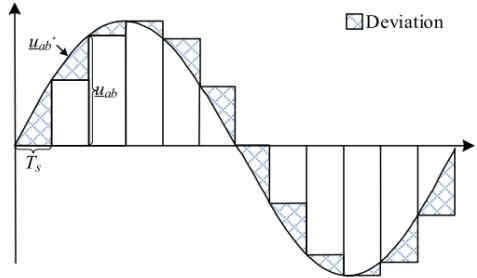

**Figure 4 - Zero Order Hold function when applied to a continuous signal [6]**

An $nth$ order data hold function will use $n+1$ previous samples of data to approximate the current value of a continuous signal [5]. Therefore, for a zero-order hold function, only the previous piece of data is used to approximate the signal. As shown in Figure 4, a zero-order hold does not generate the perfect representation of the system. The accuracy of the signal representation can be improved by increasing the order of the hold function [5]. Doing so will lead to improved signal accuracy but at the detriment of increased time delay [5], which in a closed-loop system, such as the one being developed here, can lead to significant instability being observed in the output of the control system [5].

### 2.1.2 System Discretisation
As mentioned, when we discretise the model and apply the zero-order data hold function, the input for the k$^{th}$ loop comes from the (k-1)$^{the}$ loop outputs. If we define an arbitrary coordinate $r$, that responds to an input $I$, which are both functions of time. We can then discretise the continuous function $r(t)$, as follows:

$$Let: r_1 = r(t), r_2 = \dot{r}(t) \quad (1)$$

$$r_2(k) = r_2(k-1) + \delta t \cdot f(I(k-1)) \quad (2)$$

$$r_1(k) = r_1(k-1) + \delta t \cdot r_2(k-1) \quad (3)$$

## 3 AIRCRAFT CONFIGURATION
The mathematical model, and therefore controller is based upon the previous work conducted by Wang and Cai [7].
Key Features:
1. Four rotors, which can tilt from the vertical to the horizontal position.
2. A free wing behind each rotor. The rear two free wings have a flap capable of providing pitch control

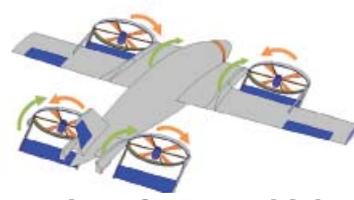

**Figure 5 - Aircraft For which controller design is based [7]**

From the paper provided by Wang and Cai [7], key dimensional data can be extracted. In addition to these variables, the additional data required for use in this paper can be summarized as:

**Table 1 - Key Dimension and Mass Variables**

| | |
|---|---|
| $l_1$ | Distance between the CoG and fore rotors |
| $l_2$ | Distance between the CoG and aft rotors |
| $l_3$ | Distance between the CoG and the center of lift for the aft free-wings |
| $l_4$ | Distance between the CoG and the center of lift fore free-wings |
| $\Delta_x$ | Distance between the fore and aft rotors |

Furthermore, variables which relate to the front rotors will have the subscript 1 and those that relate to the aft rotors will have the subscript 2. For example, $T_2$ is the combined thrust output for the rear rotors.

### 3.1 Rotor Actuators
The tilting of the rotors can be achieved in many ways, for example, in the work presented by Wang and Cai [7], rotor tilt is achieved by using one actuator and a gearbox and a series of shafts. As part of this paper, alternative methods for achieving actuation were considered.

Since the aerospace industry is largely moving towards a more electric aircraft, the actuator selected for this aircraft is the electromechanical actuator. It is essentially an electric motor (with the associated control electronics) connected to a ball screw via a gearbox [8]. Such actuators have already been used for thrust vector control and have an increased level of safety and reliability when compared to traditional hydraulic systems [8].

## 4 MATHEMATICAL MODEL
During the modelling process, the following assumptions are made:
- There is no sideslip
- The aircraft is fixed in the roll and yaw axes, I.e.: $\phi = \psi = 0$
- The rotors tilt at a constant rate ($\dot{\beta} = 8°$)
- At any given instant all rotors have an equal tilt angle ($\beta$)

Furthermore, only the continuous solutions are shown here, when the simulation is developed in section 6, these formulae are discretized using the process outlined in section 2.1.2.

## 4.1 Thrust Outputs

The forces due to the thrust of the rotors, when applied to the body frame of the aircraft can be summarised as:

$$\eta_T = T_1 \begin{bmatrix} \cos\beta \\ \sin\beta \\ l_1 \cos\beta \end{bmatrix} + T_2 \begin{bmatrix} \cos\beta \\ \sin\beta \\ -l_2 \cos\beta \end{bmatrix} \quad (4)$$

## 4.2 Fixed Wings

For the fixed-wing, it remains fixed in the horizontal position on the aircraft and as such its model is largely the same as any conventional wing. The local airspeed and the angle of attack are calculated by the usual means. This, therefore, leads to the following lift and drag forces:

$$C_{Lw} = C_{L0} + C_{L\alpha} \cdot \alpha \quad (5)$$

$$L_w = 0.5 \rho V^2 S_w C_{Lw} \quad (6)$$

$$D_w = 0.5 \rho V^2 S_w \left( C_{D0} + \frac{C_{Lw}^2}{\pi \cdot AR \cdot e} \right) \quad (7)$$

Which, when considered with reference to the body frame, results in the following set of forces:

$$\eta_w = L_w \begin{bmatrix} \cos\alpha \\ -\sin\alpha \\ 0 \end{bmatrix} - D_w \begin{bmatrix} \sin\alpha \\ \cos\alpha \\ 0 \end{bmatrix} \quad (8)$$

## 4.3 Free Wings

The free wings are located directly behind each rotor, perpendicular to the rotation plane of the rotor. For the forward free wings ($j = 1$), there is no flap bias – i.e.: the lift of the free wings is purely a function of the local airspeed and angle of attack. For the rear free wings ($j = 2$), there is an elevator flap, which is used to control the pitch of the aircraft. This, therefore, means the lift produced by the rear free wings is a function of the local airspeed, angle of attack and the flap bias angle.

Before the forces experienced by the free wings can be ascertained, the local airspeed is required to be calculated. As explained by Wang and Cai [7], the local airspeed for the free wing is a function of the translational velocities, the rotor tilt angle and the velocity induced by the rotor thrust output. The translational velocities as apparent in the rotor reference frame are calculated as shown in equation 9:

$$\begin{bmatrix} V_{\beta z} \\ V_{\beta x} \end{bmatrix} = \begin{bmatrix} -\sin\beta & \cos\beta \\ -\cos\beta & \sin\beta \end{bmatrix} \begin{bmatrix} \dot{Z} \\ \dot{X} \end{bmatrix} \quad (9)$$

From there, the rotor induced velocity (over the j'th free wing) can be calculated using the following formula:

$$v_{i,j}^4 + 2V_{\beta z} \cdot v_{i,j}^3 + \left(V_{\beta x}^2 + V_{\beta z}^2\right)v_{i,j}^2 = \frac{T_j}{2\rho\pi R^2} \quad (10)$$

Using equation 10, it is not possible to directly calculate the induced velocity. Therefore, as discussed by Wang and Cai [7], the induced velocity can be calculated using the Newton-Raphson method. For this model, the conversion of the solution is considered to be complete when the change in the estimated induced velocity, $v_{i,j}$, is less than 0.1%, i.e.: $\Delta v_{i,j} < 10^{-3}$ and $v_{i,j} \neq 0$.

Using this induced velocity, the resultant velocity over the free wing can then be calculated in the following manner:

$$V_{rj} = \sqrt{\left(V_{\beta x} + v_{i,j}\right)^2 + V_{\beta z}^2} \quad (11)$$

Knowing the translational velocities along with the induced velocity allows the local angle of attack of the free wing to be calculated. To calculate the local angle of attack the following formula can be used:

$$\alpha_{f,j} = \tan^{-1}\left[\frac{V_{\beta z}}{V_{\beta x} + v_{i,j}}\right] \quad (12)$$

Finally, the lift and drag of the free wing can be calculated conventionally, as follows:

$$C_{L,j} = C_{L0} + C_{L\alpha} \cdot \alpha_{f,j} + C_{L\delta e} \cdot \delta_{e,j} \quad (13)$$

$$L_{f,j} = 0.5 \rho V_{r,j}^2 S_f C_{L,j} \quad (14)$$

$$D_{fi} = 0.5 \rho V_{r,j}^2 S_f \left( C_{D0} + \frac{C_{L,j}^2}{\pi \cdot AR \cdot e} \right) \quad (15)$$

As before, with reference to the body frame, the forces on the aircraft can be summarised as:

$$\eta_f = \begin{bmatrix} \sin\beta \cos\alpha_{f1} \\ \cos\beta \sin\alpha_{f2} \\ l_4 \sin\beta \end{bmatrix} L_{f1} + \begin{bmatrix} \sin\beta \cos\alpha_{f2} \\ \cos\beta \sin\alpha_{f2} \\ l_3 \sin\beta \end{bmatrix} L_{f2}$$
$$+ \begin{bmatrix} \cos\beta \sin\alpha_{f1} \\ \sin\beta \cos\alpha_{f1} \\ l_4 \cos\beta \end{bmatrix} D_{f1} + \begin{bmatrix} \cos\beta \sin\alpha_{f2} \\ \sin\beta \cos\alpha_{f2} \\ l_3 \cos\beta \end{bmatrix} D_{f2} \quad (16)$$

## 4.4 Rotor Tilting Reactionary Moments

To transition from hover flight to the forward's flight regime, the rotors are required to rotate from the vertical position to the horizontal. To do this, a moment must be imparted onto the rotors. Due to the conservation of energy, any such moment will therefore be imparted onto the aircraft, just in an equal but opposite direction.

When the rotors are spinning, they effectively act in a similar manner to a gyroscope, as such when the rotors are tilted a reactionary moment, due to the tilt rate is exerted onto the aircraft. Since these rotors are being rotated around an axis parallel to the y-axis of the body frame, the reactionary moment will be around the same axis, causing a pitching moment. The magnitude of the pitching moment is governed by the rotor speed ($\Omega_j$), the moment of inertia for the rotor disc ($J_r$) and finally the angular velocity of the tilt ($\dot{\beta}$). It is known that half of the rotors spin with equal speed but in an opposite direction to the other half. Knowing, this the reactionary rotor tilting moment for a quad tilt rotor, can therefore be given as:

$$M_{\dot{\beta}} = \begin{bmatrix} 0 \\ 0 \\ 1 \end{bmatrix} \sum_{j=1}^{4} (-1)^{j+1} J_r \Omega_j \dot{\beta} \quad (17)$$

With the assumption that the rotor speed is the same for all rotors, it can also be seen that with the use of an even number of rotors, with half spinning in the opposite directions, the reactionary moment is eliminated. However, it is included here for the sake of completeness and such that the designed model and simulation can be adapted for future aircraft types.

### 4.5 Aircraft Body Frame
Summing the previously discussed elements allows the following forces (with reference to the body frame) to be defined:

$$\begin{bmatrix} F_{za} \\ F_{xa} \\ \tau_\theta \end{bmatrix} = \eta_T + \eta_w + \eta_{fw} + M_{\dot{\beta}} \quad (18)$$

### 4.6 Earth Frame
Correcting for the pitch angle and the addition of gravity, yields the following set of equations:

$$\begin{bmatrix} F_{ze} \\ F_{xe} \end{bmatrix} = \begin{bmatrix} \cos\theta & \sin\theta \\ -\sin\theta & \cos\theta \end{bmatrix} \begin{bmatrix} F_{za} \\ F_{xa} \end{bmatrix} - \begin{bmatrix} mg \\ 0 \end{bmatrix} \quad (19)$$

### 4.7 Resulting Accelerations
Applying Newton's second law yields:

$$\begin{bmatrix} \ddot{Z} \\ \ddot{X} \\ \ddot{\theta} \end{bmatrix} = \begin{bmatrix} m^{-1} & 0 & 0 \\ 0 & m^{-1} & 0 \\ 0 & 0 & J_y^{-1} \end{bmatrix} \begin{bmatrix} F_{ze} \\ F_{xe} \\ \tau_\theta \end{bmatrix} \quad (20)$$

### 4.8 Measurement Noise
Any measurements taken on the aircraft will have an error in the measurement, simply due to the vibrations on the aircraft and the limits of the measurement equipment. This error will manifest itself as "white" noise and will essentially be random. Since there are a sufficient number of data points, it can be assumed that any noise will be normally distributed.

To simulate the random noise generation, the MATLAB function, randn, is used. This function will generate a number at random between $\pm 5$. It would be seemingly impossible for a controller to reject noise of this magnitude ($\pm 5m$). Therefore, the proportionality constant κ, is used for each degree of freedom. As such the noise for each degree of freedom is generated as below:

$$N_j = \kappa_j \cdot randn(1,1) \quad (21)$$

In the model, this noise will be summated with the controlled variables (i.e.: $Z, \dot{X}, \theta$), to simulate the presence of noise in system measurements.

## 5 CONTROL SYSTEM DESIGN

### 5.1 Measurement and Filtering
The presence of noise in a control system is undesirable, as it can lead to system instability. Therefore, before adequate control of the system can be ascertained, it is required as much noise is rejected from the system as possible. The use of the filter helps to ensure the noise present in the system is rejected, ensuring closed-loop stability and robust performance with regard to set point tracking [9].

Since the modelled noise is high frequency in nature, a low pass filter of the first order is selected. This allows signals with frequencies below the cut-off frequency ($\omega_0$), to pass through the filter with all others being rejected. In the digital domain, the first order filter for a generalised coordinate $r$, has the following form:

$$r_f(k) = \frac{1}{1 + \delta t \cdot \omega_0} \left( r_f(k-1) + \omega_0 \cdot r_m(k) \right) \quad (22)$$

Where possible it is beneficial to minimize the cut off frequency, as selecting a value that is too stringent leads to a delay in the change of the signal being observed, which as previously discussed, significant delays can ultimately lead to instabilities being observed in the control system.

### 5.2 Generalised Controller
The behaviour of a digital PID controller is largely the same as that in the continuous form. To convert the PID controller to the digital domain, the data hold function discussed in section 2.1.1, must be applied. This, leads to the error in the digital domain to be defined as:

$$e_r(k) = r_d(k) - r(k) \quad (23)$$

$$\dot{e}_r(k) = \delta t^{-1} \cdot \big(e(k) - e(k-1)\big) \quad (24)$$

$$\int e_r(k) = \sum_{n=0}^{k} (e(n) \cdot \delta t) \quad (25)$$

From this, the digital PID controller can be defined in the following manner:

$$\ddot{r}(k) = k_{pr} e_r(k) + k_{ir} \int e_r(k) dt + k_{dr} \dot{e}_r(k) \quad (26)$$

$$u_r(k) = m_r \ddot{r}(k) \quad (27)$$

## 5.3 UAV Hover Mode

In the hover mode, we seek to control all three degrees of freedom. In this mode the airspeed of the aircraft is approximately zero, this, therefore, means the lift produced by the main and free wings is negligible. This, therefore, means the dynamics of the system are controlled directly by the thrust output of each rotor.

### 5.3.1 Altitude Control

In the z-direction, the altitude is controlled by the total thrust in the z-direction, which is a function of the total thrust output and the pitch and tilt angles. From this, we can define the total thrust output as:

$$T_T(k) = \frac{u_z(k)}{\cos\theta(k)\cos\beta(k)} \quad (28)$$

### 5.3.2 X-Velocity Control

Furthermore, in this mode, the x-velocity is controlled by the magnitude of the thrust in the x-direction. Since the rotors are approximately vertical, they provide minimal thrust in the x-direction and as such, to control the x-velocity, a component of the total thrust output must be redirected in the horizontal direction. To do this, we select the pitch angle to control the x-velocity. The desired pitch angle can therefore be selected as follows:

$$\theta_d(k) = \sin^{-1}\left(\frac{-u_x(k)}{T_T(k)}\right) \quad (29)$$

### 5.3.3 Pitch Control

As mentioned, the flight speeds during this phase of flight are small, and as such the pitch moment is entirely controlled by creating a moment about the y-axis, in this case, this moment is generated by the differential in thrust between the two rotors. The thrust differential between the two rotors is defined as follows:

$$u_\theta(k) = [T_1(k) l_1 - T_2(k) l_2] \cos\beta(k) \quad (30)$$

### 5.3.4 Combined Control Output Matrix

By combining the three control outputs as defined in equations 28, 29, 30, The control output for each rotor and the elevator flap can be selected as:

$$\begin{bmatrix} T_1(k) \\ T_2(k) \\ \delta_e(k) \end{bmatrix} = \begin{bmatrix} T_T - \frac{1}{\Delta_x}\left\{T_T(k) l_1 - \frac{u_\theta(k)}{\cos\beta(k)}\right\} \\ \frac{1}{\Delta_x}\left\{T_T(k) l_1 - \frac{u_\theta(k)}{\cos\beta(k)}\right\} \\ 0 \end{bmatrix} \quad (31)$$

## 5.4 Control for mode transition

This flight mode is the most challenging to control as, in this mode, the vehicle dynamics change drastically. At the start of this phase, the airspeeds are low and as such the control torque available from the free wings is minimal, however, the magnitude of the torque available from the rotors is significant. As the rotors tilt to the horizontal position, the airspeed of the vehicle naturally increases. As the airspeed increases, the amount of torque available from the free wings is significant but because of the large tilt angles, the torque available from the rotors is decreased.

### 5.4.1 Altitude Control

For the majority of this flight mode, the airspeed is below the stall speed of the aircraft, meaning the altitude of the aircraft must be controlled by altering the total thrust acting in the vertical direction. Therefore, the total rotor thrust can be selected using the following formula:

$$T_T(k) = \frac{u_z(k)}{\cos\theta(k)\cos\beta(k)} \quad (32)$$

### 5.4.2 X-Velocity Control

During this flight stage, the x-velocity is not controlled. The increase in x-velocity is achieved simply due to the rotors tilting to the horizontal position.

### 5.4.3 Pitch Control

Due to the absence of x-control, the desired pitch angle is selected using qualitative and quantitative analysis of the simulation, with adjustments made until satisfactory performance over the entire flight plan is ascertained.

In this mode, pitch control is achieved from two sources, one from the flight control surfaces (free wings) and the other from the rotors. During the rotor tilting, the local angle of attack to the free wings increases significantly. Due to this the lift generated by the wing increases drastically which therefore causes a significant pitching moment to be generated. To combat this and thus minimize the effect on the rotor thrust, the elevator flap deflection is selected such that the moment between the fore and aft free wings is zero. This, therefore, means the elevator flap bias angle is selected as follows:

$$\delta_{e(m=0)}(k) = \frac{1}{l_3 \rho V_{r2}^2(k) S_f C_{L\delta e} \sin \beta(k)}$$
$$\cdot \left[ l_4 \left( L_{f1}(k) \sin \beta(k) + D_{f1}(k) \cos \beta(k) \right) \right.$$
$$- l_3 \left( \rho V_{r2}^2(k) S_f C_{L\alpha} \alpha(k) \sin \beta(k) \right.$$
$$\left. + D_{f2}(k) \cos \beta(k) \right) \right] \quad (33)$$

The rotors are used to overcome the rest of the disturbances and as such, the thrust differential between the two rotors can again be defined in the same manner as the hover controller, as follows:

$$u_\theta(k) = [T_1(k) l_1 - T_2(k) l_2] \cos \beta(k) \quad (34)$$

### 5.4.4 Combined Control Output Matrix
By combining the selected control outputs as defined in equations 32, 33, 34, the control outputs for each set of rotors and the elevator flap can be specified as:

$$\begin{bmatrix} T_1(k) \\ T_2(k) \\ \delta_e(k) \end{bmatrix} = \begin{bmatrix} T_T - \frac{1}{\Delta_x} \left\{ T_T(k) l_1 - \frac{u_\theta(k)}{\cos \beta(k)} \right\} \\ \frac{1}{\Delta_x} \left\{ T_T(k) l_1 - \frac{u_\theta(k)}{\cos \beta(k)} \right\} \\ \delta_{e(m=0)} \end{bmatrix} \quad (35)$$

## 5.5 Forwards Flight
In this flight regime, the airspeeds are relatively high and as such, the flight control surfaces, and the wing provides the majority of the lift force and control torque.

### 5.5.1 Altitude Control
The altitude of the aircraft is controlled by the total amount of force in the z-direction, the major contributor to this force is largely the lift produced by the wing. The lift produced by the wing is proportional to the angle of attack experienced by the wing. It is not possible to directly augment the angle of attack. Therefore a pitch angle ($\theta$) that will yield the desired angle of attack once the climb angle ($\gamma$) is considered must be selected. The desired values of angle of attack can be selected as:

$$\alpha_d(k) = [u_z(k) - T_T(k) \sin \theta(k) \sin \beta(k)] \cdot [q(k) S_w C_{L\alpha} \cdot \alpha(k)]^{-1} \quad (36)$$

This, therefore, yields the following desired pitch angle as:

$$\theta_d(k) = \alpha_d(k) + \gamma(k) \quad (37)$$

### 5.5.2 X-Velocity Control
In this flight mode, the rotors are approximately horizontal (i.e.: $\beta \approx 90°$), this, therefore, means the thrust output from the rotors directly controls the x-velocity. Therefore, selecting the total thrust output is trivial and is therefore selected using the following means:

$$T_T(k) = \frac{u_x(k)}{\cos \theta(k) \sin \beta(k)} \quad (38)$$

### 5.5.3 Pitch Control
Furthermore, we know the pitch of the aircraft is controlled by the moment generated by the free wings. We also know that the forward free wings are fixed in their position and geometry. This, therefore, means the only way we can generate a control torque is to augment the lift of the rear free wing. This is done by moving the elevator flap. The deflection of this flap can be selected using the following formula:

$$\delta_{ec}(k) = \frac{1}{l_3 \rho V_{r2}^2(k) S_f C_{L\delta e} \sin \beta(k)}$$
$$\cdot \left[ l_4 \left( L_{f1}(k) \sin \beta(k) + D_{f1}(k) \cos \beta(k) \right) \right.$$
$$- l_3 \left( \rho V_{r2}^2(k) S_f C_{L\alpha} \alpha(k) \sin \beta(k) \right.$$
$$\left. + D_{f2}(k) \cos \beta(k) \right) + u_\theta(k) \right] \quad (39)$$

### 5.5.4 Combined Control Output Matrix
By combining equations 36, 37, 38, 39, the control outputs of the system are then defined as follows:

$$\begin{bmatrix} T_1(k) \\ T_2(k) \\ \delta_e(k) \end{bmatrix} = \begin{bmatrix} 0.5 T_T(k) \\ 0.5 T_T(k) \\ \delta_{ec}(k) \end{bmatrix} \quad (40)$$

## 5.6 Control Switching Logic
As discussed, the three different flight states have three different control implementations. As shown in the previous sections, these control implementations are distinct and their effectiveness is dependent on the tilt angle of the rotors and the aircraft's airspeed. This, therefore, means it is important to establish when each flight mode begins and ends.

The hover controller is enabled when the rotors are in the vertical position, i.e.: $\beta = 0°$. When the transition is commanded, the rotors begin to tilt, and the aircraft begins to accelerate in the x-direction. The transition controller remains enabled until the airspeed of the vehicle exceeds 57m/s, which equates to a tilt of $\beta \approx 80°$, at which point there is sufficient airspeed for altitude and controllability to be maintained. When these values are exceeded the forward's flight controller is enabled.

## 5.7 Selected Control Parameters

As discussed in the previous section, the three different flight modes, have three different control laws. As a result of this, the effects of the PID controller and any apparent system noise will be different for each flight mode. For these reasons, it becomes necessary to alter the control gains and the magnitude of filtering such that stability can be maintained.

The control parameters for each flight mode are outlined in the table below. These have been selected by using qualitative analysis of the results obtained in the various flight mode simulations.

**Table 2 – Selected Control Gains and Filter Cut-Off Frequencies**

| Parameter | Hover | Transition | Forwards |
|---|---|---|---|
| $k_{pz}$ | 2 | 2.4 | 0.94 |
| $k_{iz}$ | 0.05 | 0.02 | 0.03 |
| $k_{dz}$ | 8 | 3 | 3 |
| $k_{px}$ | 0.6 | - | 1.4 |
| $k_{ix}$ | 0.02 | - | 0.3 |
| $k_{dx}$ | 2 | - | 2 |
| $k_{p\theta}$ | 7 | 9 | 1.75 |
| $k_{i\theta}$ | 0.5 | 0.085 | 0.01 |
| $k_{d\theta}$ | 7 | 9 | 2.5 |
| $\omega_z$ | $3\ s^{-1}$ | $3\ s^{-1}$ | $3.25\ s^{-1}$ |
| $\omega_x$ | $3.5\ s^{-1}$ | - | $1\ s^{-1}$ |
| $\omega_\theta$ | $1\ s^{-1}$ | $1\ s^{-1}$ | $6.2\ s^{-1}$ |

## 6 SIMULATION

As discussed in order to demonstrate that the designed controller can maintain stability throughout the three different modes the discussed model and digital controller have been implemented in the MATLAB software suite. Two conversion cases are considered, hover to forwards flight and then forwards flight to hover. The simulation runs over $250s$, with the time step selected as $\delta t = 0.1s$.

In the first half of the simulation, take-off with followed by hover to forwards flight transition is considered. This process can be described as follows: The aircraft starts at rest, at the home position, with all displacements and velocities of interest set to zero $(Z, \dot{X}, \theta) = (0,0,0)$. The aircraft will then take-off, and will move to the desired altitude of $(Z_d = 50m)$. Once this is complete, the aircraft will then transition to the forwards flight mode. Whilst in the forwards flight mode, the controller will then seek to maintain the desired altitude $(Z_d = 50m)$ and desired translational velocity $(\dot{X}_d = 57ms^{-1})$.

The second half of the simulation focuses on the transition between forwards flight and hover and then achieving a controlled landing. During this phase, the aircraft will begin to transition from the forwards mode back to the hover mode. Due to high airspeeds and therefore large forces acting on the UAV the transition from forwards to hover flight is not straight forwards and as such an alternate protocol had to be developed such that the airspeed can be sufficiently reduced. A full breakdown of the selected reverse-transition protocol can be seen in Figure 6.a.

## 7 ANALYSIS OF RESULTS

From the results shown in Figure 6, it can be seen that the control challenges have largely been overcome. However, the quality achieved by the controllers of the steady-state processes (Hover, Forwards Flight) are significantly superior to those achieved by the controller of the dynamic processes (Transition).

### 7.1 Hover Mode

As shown in Figure 6, the implementation of the hover controller is successful, with the hover controller maintaining stability in all three of the controlled degrees of freedom. The performance of the controller is as expected, with noise being successfully rejected from the system and the error between the desired position and the actual position tending to zero. For the hover controller, there is negligible overshoot in the z-direction. The x velocity is successfully controlled with all disturbances removed. The pitch controller is successfully implemented, with the measured value of the system closely matching that requested by the x-velocity controller. In terms of the thrust output and the desired pitch angle, a small amount of what looks like noise remains in the system. This is due to the summing of all the noise from all three degrees of freedom, the remaining measurement noise and control output requests then being summated together and directly selecting each rotor thrust and desired pitch angle.

### 7.2 Forwards Flight Mode

As shown in the figures, the forward's flight control is also implemented successfully. Apart from the initial overshoot as the transition mode becomes forwards flight and the slight control ripple, the altitude is successfully controlled, with the error tending to zero with the desired value being achieved. Again, for the x-velocity, apart from the initial ripple during mode change, the x-velocity is successfully controlled, with the noise being rejected and the overall system error tending to zero. From the figures presented it can be seen that the thrust required in the forwards flight mode, is (on average) less than that of the hover and transition modes. This further demonstrates the advantages a tilt rotor aircraft has over a conventional rotorcraft. With lower thrust requirements meaning lower fuel burn for the same cruise speed, therefore increasing range and endurance.

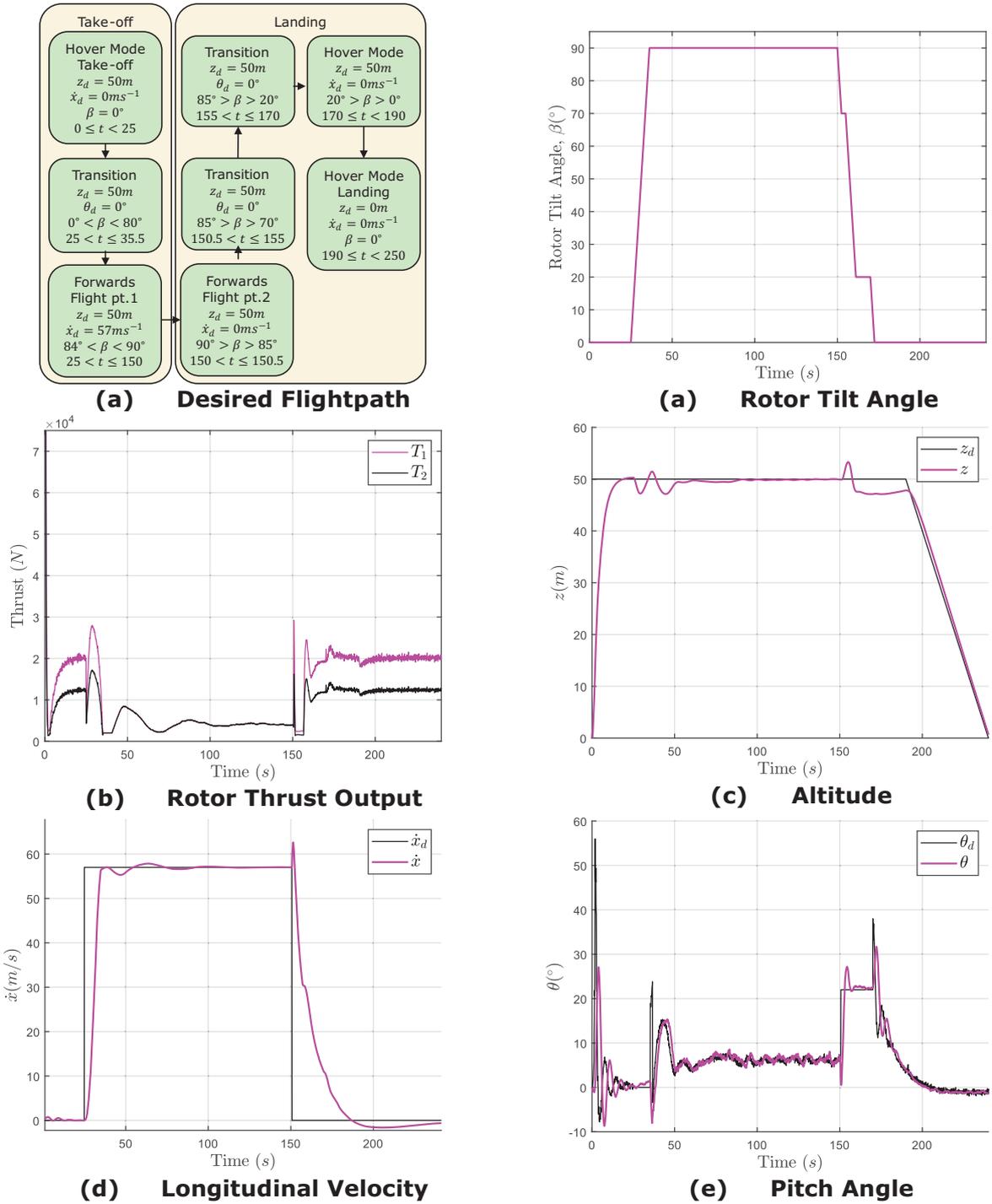

Figure 6 - Flight Mode Transition Control for Tilt Rotor UAV (Hover → Flight → Hover)

### 7.3 Mode Transition

For both the transition phases, it can be seen that the mode transition is completed successfully. With the key challenges (altitude and attitude control) largely being addressed and overcome. Success with respect to these challenges can be seen by the fact the altitude and attitude remaining stable, with only small divergences from the setpoint.

For both transition phases (hover to forwards and forwards to hover), it can be seen the pitch tends to the set point value, with a small amount of overshoot and subsequent ripple being observed. These divergences could be removed with further tuning of the controllers and a more sophisticated control division algorithm (such as the one discussed in section 8.2) being implemented. Furthermore, it can be seen that that for both transition phases, there is a small amount of altitude deviation, approximately $\pm 4m$ in magnitude. Whilst not a major issue when at altitude it would be preferable for this deviation to be removed.

Although a significantly different control algorithm and schema are used in the paper by Wang and Cai [7], similar levels of divergence in the altitude are also observed in the transition phase. This, therefore, suggests that further refinements with regards to the transition protocol needs to be carried out. Potentially, a more sophisticated control algorithm may need to be implemented to ensure the altitude is completely invariant.

## 8 FURTHER WORK
As mentioned, whilst the design of this PID controller is successful, in that all three flight modes can be controlled successfully, a more sophisticated control algorithm is likely required before implementing the presented PID controller in the real world.

### 8.1 Integrator Wind-up
It is possible that during some flight manoeuvres the desired control outputs exceed the physical limits of the actuator. When this happens, the feedback loop to the PID controller is essentially broken [10]. As a result, the integral term will begin to increase in value significantly, which once the error changes sign, will lead to significant overshoot of the desired value [10].

This paper uses set point limitation to prevent the actuators from becoming saturated, however, this leads to conservative bounds on the system performance and does not prevent saturation due to disturbances [10].

A more sophisticated method, such as Back-Calculation and Tracking needs to be implemented. When saturation of the thruster occurs, back-calculation recomputes the integral term such that the controller output is at the maximum or minimum of the "actuator" [10]. By using such a system, the integral term would not wind up significantly, leading to the superior performance of the system [10].

### 8.2 Control Division (Allocation)
Traditionally, the control of an aeroplane has been achieved by allocating one effector to each rotational degree of freedom. For this aircraft, the pitching dynamics in the transition mode can be controlled by two effectors, the rotor thrust and the elevator deflection. In the work presented in this paper, the allocation of the control torque has been rudimentary at best, with the elevator deflection being selected such that the aerodynamic torque is negligible. Therefore, one of the key challenges with the implementation of the transition controller has been the control allocation i.e.: how much of the pitching moment control is allocated to the rotor-thrust differential and the elevator.

Control allocation algorithms aim to find the "best fitted" allocation of effector outputs [11]. Such algorithms can also be used to achieve secondary goals, such as power and drag minimisation [11]. The application of a sophisticated control allocation to replace the rudimentary one placed here will add additional complexity to the controller but should be able to improve the performance (particularly in the transition phase).

## 9 CONCLUSION
This paper sought to model and design a digital control system for the longitudinal dynamics of a preconceived conceptual quad tilt-rotor UAV. A model and controllers for all three flight modes has been designed along with all control gains and filtering parameters being selected. As discussed, the performance of the controllers could be improved, along with some additional improvements to robustness via integrator anti-windup and the use of control allocation algorithms. However, the results outlined in this paper demonstrate that the simple digital PID controller can be successfully used to reject noise and undesired aerodynamic forces, and therefore allow stable control of the aircraft in all three flight modes.